\documentclass[12pt]{iopart}

\usepackage{iopams}  
\usepackage{graphicx}
\usepackage{xcolor}

\newcommand{\noi}{\noindent}
\newcommand{\be}{\begin{equation}}
\newcommand{\ee}{\end{equation}}

\begin{document}

\title[Hierarchical invasion of cooperation in complex networks]{Hierarchical invasion of cooperation in complex networks}

\author{Daniele Vilone}
\address{LABSS (Laboratory of Agent Based Social Simulation), Institute of Cognitive Science and Technology, National Research Council (CNR), Via Palestro 32, 00185 Rome, Italy}
\address{Grupo Interdisciplinar de Sistemas Complejos (GISC), Departamento de Matem\'aticas, Universidad Carlos III de Madrid, 28911 Legan\'es, Spain}
\ead{daniele.vilone@gmail.com}
\author{Valerio Capraro}
\address{Center for Mathematics and Computer Science (CWI), 1098 XG, Amsterdam, The Netherlands}
\address{Business School, Middlesex University London, NW44BT, London, United Kingdom}
\author{Jos\'e J. Ramasco}
\address{Instituto de F\'{\i}sica Interdisciplinar y Sistemas Complejos IFISC (CSIC-UIB), 07122 Palma de Mallorca, Spain}

\begin{abstract}
The emergence and survival of cooperation is one of the hardest problems still open in science. Several factors such as the existence of punishment, repeated interactions, topological effects and the formation of prestige may all contribute to explain the counter-intuitive prevalence of cooperation in natural and social systems. The characteristics of the interaction networks have been also signaled as an element favoring the persistence of cooperators. Here we consider the invasion dynamics of cooperative behaviors in complex topologies. The invasion of a heterogeneous network fully occupied by defectors is performed starting from nodes with a given number of connections (degree) $k_0$. The system is then evolved within a Prisoner's Dilemma game and the outcome is analyzed as a function of $k_0$ and the degree $k$ of the nodes adopting cooperation. Carried out using both numerical and analytical approaches, our results show that the invasion proceeds following preferentially a hierarchical order in the nodes from those with higher degree to those with lower degree. However, the invasion of cooperation will succeed only when the initial cooperators are numerous enough to form a cluster from which cooperation can spread. This implies that the initial condition has to be a suitable equilibrium between high degree and high numerosity. These findings have potential applications to the problem of promoting pro-social behaviors in complex networks.

\end{abstract}

%Uncomment for PACS numbers title message
%\pacs{00.00, 20.00, 42.10}
% Keywords required only for MST, PB, PMB, PM, JOA, JOB? 
%\vspace{2pc}
%\noindent{\it Keywords}: Article preparation, IOP journals
% Uncomment for Submitted to journal title message
%\submitto{\JPA}
% Comment out if separate title page not required
\maketitle

\section{Introduction}

How cooperation surges and becomes stable despite the tension introduced by individual interest is one of the most debated questions across sciences~\cite{boyd2003,fehr2003,santos2005,nowak2006,perc2010,rand2013,capraro2013,dorsogna2015}. Individual interest implies the search for the own outcome optimization, although it usually leads to sub-optimal solutions at a community or global scale. Cooperation, on the contrary, may bring better global results but it requires individuals to relinquish part of their benefits to others. When the game payoff is attached to fitness in evolutionary game theory, only individually optimal strategies proliferate and cooperative behaviors are thus doomed to disappear in a few generations. Such grim expectations are challenged by the widespread presence of cooperation in human ~\cite{frey2004,traulsen2010,apicella2012,capraro2014,capraro2014b} and animal societies~\cite{brosnan2002,clutton-brock2009,nowak2010,lukas2012}. Furthermore, cooperation is at the basis of multicellular organisms~\cite{smith1995,sachs2011,nadell2013}. All these examples occur despite the presence of strong individual incentives to default on the group cooperation. There exist, of course, counterexamples where the non-collaborative strategies dominate such as criminal activity that may be seen as non-cooperative behavior within human societies, or the loss of growth control exhibited by cancer cells within biological organisms. The key question of which factors favor the proliferation and eventual generalization of cooperation thus remains open.    

Several mechanisms have been advanced for explaining the persistence of cooperation. Some of them include various procedures for punishing free riders~\cite{boyd1992,fehr2000,fehr2002,gurerk2006,szo11,szo13b,che14,ohd16}, rewarding cooperators~\cite{milinski2002,panchanathan2004,milinski2006,capraro2016partner}, or a combination of both~\cite{andreoni2003,rockenbach2006,sefton2007,hilbe2010}, which effectively change the payoff balance. Others consider repeated interactions and the possible development of prestige \cite{milinski2002,capraro2016partner,axelrod1984,giardini2016}. When the agents have to play many times together, the inclination to cooperate may enhance if both parts benefit in long term and a trust relation can be built. Even though this can only be an explanation in some particular contexts, finite size fluctuations can also lead to the invasion and fixation of a disadvantageous strategy \cite{nowak2004}. The structure of the interaction networks have been also claimed to play a role in increasing global cooperation levels \cite{perc2010,pacheco2006,santos2006,gomez2007,szabo2007,santos2008}. Recent empirical and theoretical results, however, show that this effect may be in doubt for social systems \cite{traulsen2010,grujic2010,suri2010,gracia2012,gracia2012b,gargiulo2012}.

Here we take a different perspective. Instead of on a final stationary state, the focus is set on the invasion process of cooperators in a finite population initially almost full of defectors. As explained before, different factors may lead to the fixation of cooperation, but how does this process take place? The question that we address here is whether the structure of the interaction networks can influence the dynamics of invasion of the system by cooperative strategies.
In particular, while studies about the probability for cooperators to invade effectively a system or to survive in a hostile environment have been already accomplished~\cite{santos2006,szabo2007,lie05,oht06}, in this work we aim to understand the precise dynamical process through which the invasion takes place, if and when it does.
For this, we set initially the population in a defection state except for a few agents, and then checked the evolution of these invaders. The spatial interactions between elements of the system are modeled by two types of random networks in which the nodes are the agents and the interactions links: scale-free (configurational model) and Erd{\H o}s-R\'enyi. The invasion process is analyzed as a function of the degree of the initial cooperative nodes. The strategic interactions between agents are modeled by a Prisoner's Dilemma, arguably one the most extensively studied game-theoretical models of human cooperation~\cite{lie05,oht06,now92,roc09}. In this work we do not add further ingredients to the classical Prisoner's Dilemma Game, as for instance sanctioning defectors by cooperators~\cite{szo13,che15,per15}, reputation~\cite{gia16}, or emotions and complex internal dynamics of agents~\cite{gia15,ohd17a,ohd17b}, because here we aim to single out the network effects on the invasion process; we only vary the update algorithm, as we explain in the following, in order to test the robustness of the results we found.
We find numerically and analytically that, in this context, the invasion of cooperation follows a clear pathway passing from nodes with high to those with low degrees. This mechanism strongly mediates the invasion process and its final outcome. 

\begin{figure} 
\centering
\includegraphics[width=8.6cm]{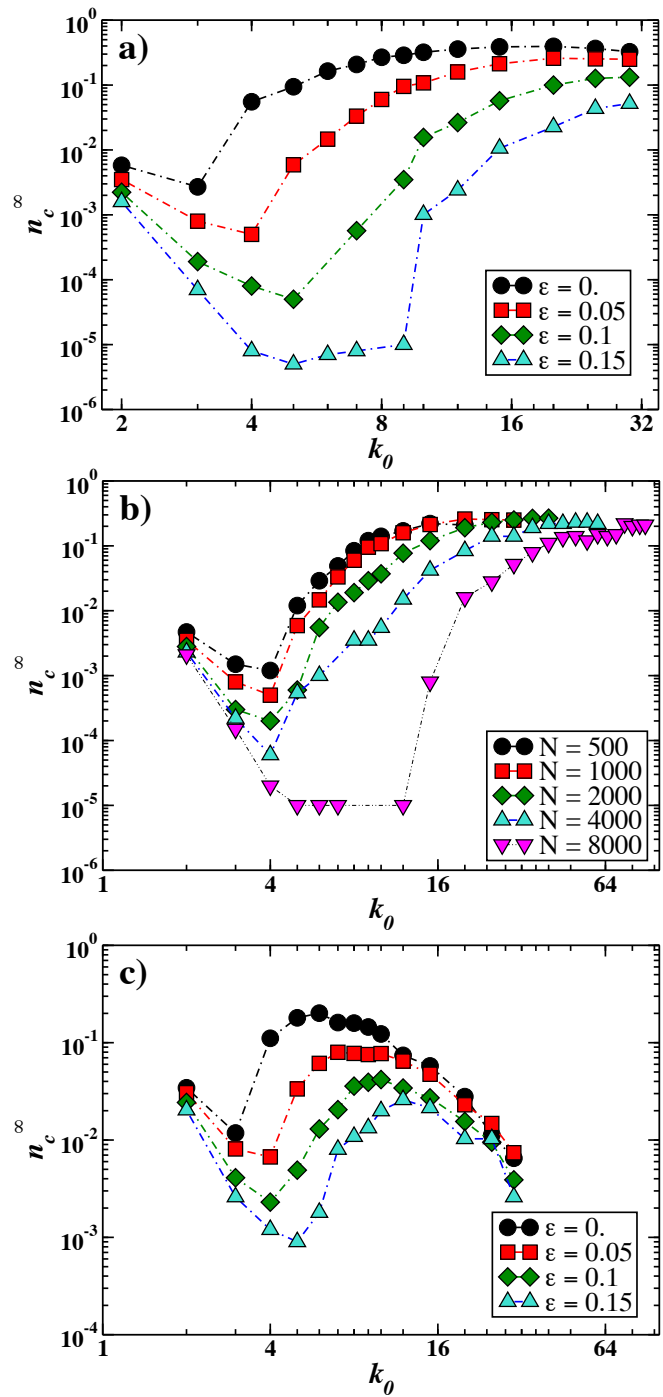} 
\caption{\label{coopSF} Final cooperator density as a function of the invasion degree $k_0$ for a system on a scale free network. In {\bf a)}, exponent $\beta=1.6$, size $N=1000$ and four different values of the punishment; In {\bf b)}, exponent $\beta=1.6$, $\varepsilon=0.05$, and five different system sizes; In {\bf c)}, exponent $\beta=2.7$, size $N=1000$ and four different values of the punishment.}
\end{figure}

\      

%%%%%%%%%%%%%%%%%%%%%%%
\section{The model}
\label{themodel}

We consider a system constituted by $N$ agents occupying the nodes of a given network. Each agent interacts directly only with her nearest neighbors, and can adopt two possible strategies:
cooperation (C) or defection (D). The interactions are given by
a Prisoner's Dilemma game, in which nodes play with their neighbors and collect a
payoff according to the action adopted by themselves and their
opponents. Payoffs are collected according to the following
matrix~\cite{vil14}:

\be
\hat{P}\ \ \ = \ \ \ 
\begin{tabular}{|c|c|c|}
\hline
\mbox{ } & {\bf C} & {\bf D} 
 \\ \hline
{\bf C} & $1$ & $0$
 \\ \hline
{\bf D} & $1.4$ & $\varepsilon$ \\
\hline 
\end{tabular} \, ,
\label{3b} 
\ee

\noi where the punishment parameter $\varepsilon$ must fall in the interval $[0,1)$.
Among all the possible and equivalent shapes, we have chosen this kind of payoff matrix because it allows us to explore what happens in all the (weak) Prisoner's Dilemma range: referring to the classical parametrization~\cite{santos2005,roc09}, we set $T=1.4$ (temptation), $R=1$ (reward), $P=\varepsilon$ (punishment) and $S=0$ (sucker's payoff); the chosen value of $T$, already used in literature~\cite{vil14}, has the advantage that is larger than the difference $P-S$, so that a defector gains more fitness with respect to the opponent when plays against a cooperator than another defector. More precisely, the dynamics takes place as follows: at each elementary time step an agent $i$, picked up at random, plays a round of the game with her neighbors. After this, each of her neighbors play a round of the game with their neighbors. Subsequently, the agent $i$ imitates the strategy of the most successful neighbor provided that her own payoff is lower. Otherwise, nothing happens. This way of updating the strategies is the so-called Unconditional Imitation (UI) rule~\cite{roc09,nowak1993,nowak1994} and it ensures that the most successful strategy rapidly spreads across the population. After that, the payoff of the players is set to zero, so that every evolution act takes place only on the basis of the last round of  the game. A time unit is made up by $N$ elementary steps of the dynamics (Monte Carlo steps). We stress the fact that at each round the fitness of a player is the simple sum of the payoffs she has collected with all her neighbors, without any normalization with respect to her degree: this choice, which is largely present in literature~\cite{oht06,now92,roc09,ich17,per17}, is actually realistic, since in nature and human communities having more interactions with peers entails generally higher fitness, being actually a possible mechanism enhancing cooperation; indeed, in real situations the agents (be they humans or not) do not compute an average, but act according to their actual fitness, since it is usually very hard in human societies, and practically impossible in nature, to get reliable information about the fitness gained by others~\cite{smith1995,roc09,footn}. Nevertheless, for sake of completeness, in Sec.~\ref{av_p} we will consider the case where the payoff collected by a player is divided by her degree.

Even though there are many possible game-theoretical models useful for describing cooperation phenomena, we consider the Prisoner's Dilemma game for two main reasons: firstly, out of the possible two-player games the conflict between the individual and group utilities is the greatest~\cite{now92,roc09}, so the mechanisms which foster cooperation can be efficiently identified. On the other hand, the dynamics of this model has been extensively studied in the literature~\cite{lie05,oht06,roc09}. This sets a baseline against which we can compare our results; moreover, it opens the door to the analysis of the effects of the update rule in the invasion process, given that the update rules and their impact on the final outcome have already been deeply studied in the context of the Prisoner's dilemma.

Contrariwise, the use of the update rule is delicate because under some conditions different rules may yield diverging results, so that it is always important to check if the results are robust by changing the evolution algorithm and inserting stochasticity in it~\cite{roc09}. In this case, we select UI as the first option for the sake of simplicity but we have checked that the same invasion patterns are observed with other updating rules. In particular, we have used the replicator (REP) update, in which after each game round the evolving individual $i$ imitates the strategy of a randomly selected neighbor with probability proportional to the payoff difference between them provided that the neighbor's payoff is higher than $i$'s~\cite{schuster1983,nowak2004b,schlag1998}. Besides UI and REP updates, we have also explored more realistic rules such as the so-called moody cooperation inspired by the findings in experimental settings \cite{gracia2012b,gru14}. In this rule the probability of modifying a strategy depends on the success or not of the last game round and on the previous strategy of the agent. In all cases, we have found similar results in the direction of the invasion (top-down) and in the characteristics that initial invaders should have so that the invasion may succeed (high degree and numerosity).

{\it Networks} - In this work we tested the behavior of the model in different topologies.
In particular, we utilized Erd{\H o}s-R\'enyi (ER)~\cite{erd60} and Scale Free (SF) random networks generated by the
Molloy-Reed algorithm~\cite{mol95}. The main difference between these types of random networks is the
heterogeneity in the number of nodes' connections (degree, $k$). In the case of ER graphs the degree distribution is Poissonian with a given average degree $\langle k \rangle$, while in the Molloy-Reed networks it decays as a power-law with an exponent $\beta$ ($P(k) \sim k^{-\beta}$). As a consequence, the number of high degree nodes is much larger in the SF than in ER networks.
  
{\it Initial conditions} -   Since our aim is to study if and how cooperation invade a system of interacting
individuals, we consider systems where initially all the agents are defectors, apart from the ones
occupying nodes of a given degree $k_0$. In this case, we consider two options: either all the agents in nodes with $k_0$ are initially cooperators or, in Sec.~\ref{fixed_n}, only a certain fixed number of nodes with degree $k_0$, $N_c^0$, selected at random are cooperators, the rest are defectors. The results were obtained exploring different
values of $k_0$.

\begin{figure}
\centering
\includegraphics[width=8.6cm]{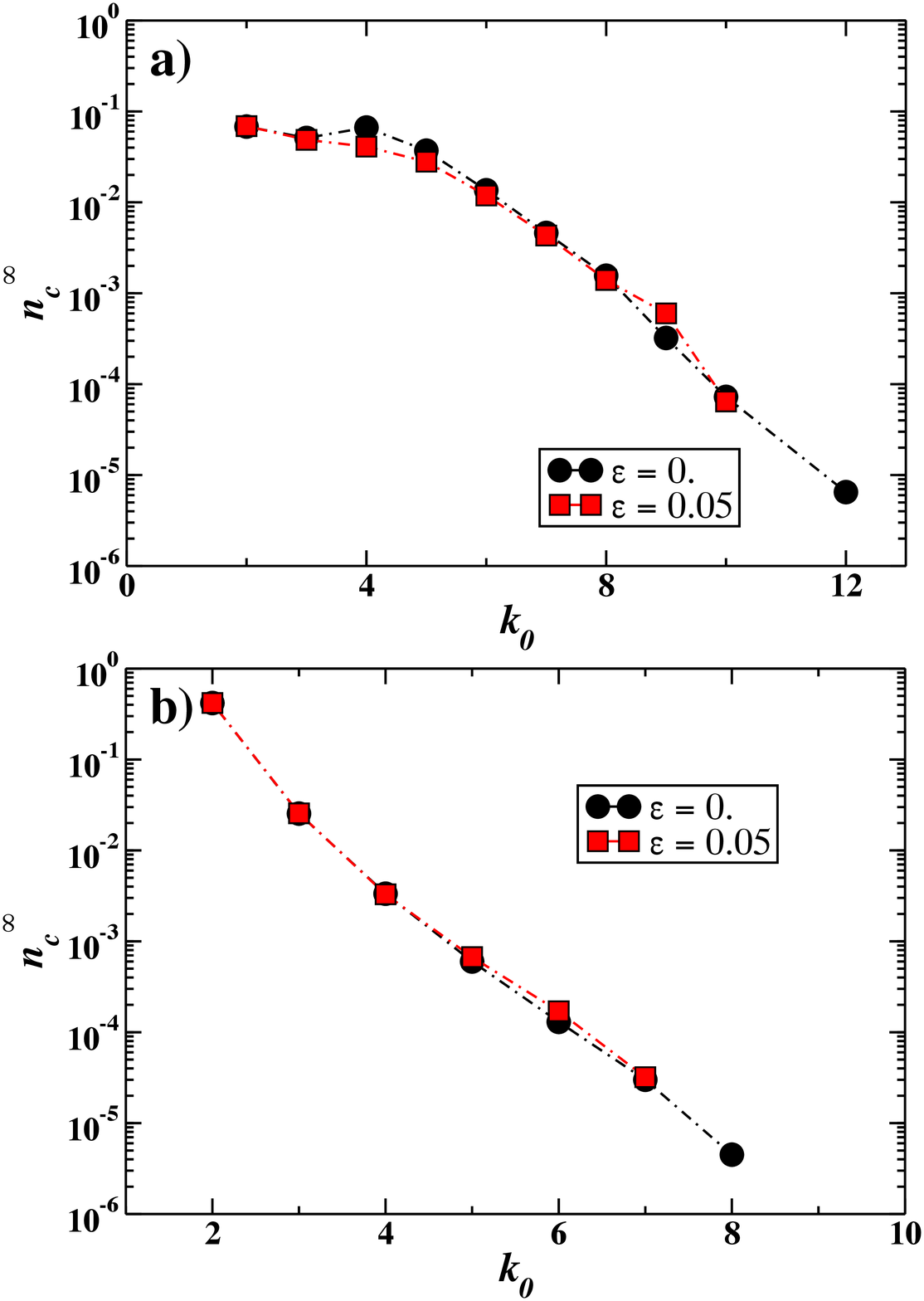} 
\caption{\label{coopER} Final cooperator density as a function of the invasion degree $k_0$ for a system on an ER network in {\bf a)} with average degree $\langle k\rangle=3.5$, size $N=1000$) and two different values of the punishment. In {\bf b)}, the same results on a scale free network with exponent $\beta=5.6$ and size $N=1000$), and two different values of the punishment.}
\end{figure}

\ 

\section{Numerical results}
  
\subsection{Final state of the system}
  
We begin the description of the results by analyzing the final configuration of the system using numerical simulations and as a function of the invasion degree $k_0$ for different values of the remaining parameters. If not explicitly specified, the system size is set at $N=1000$, although several system sizes have been explored. The results are always averaged over $2000$ independent realizations.

\begin{figure}
\centering
\includegraphics[width=8.6cm]{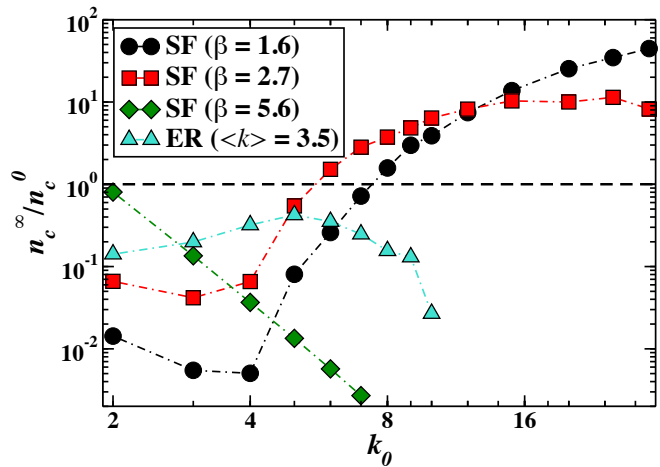}
\caption{\label{glob_c} Values of ratio between the final cooperator density $n_c^{\infty}$ over the initial one $n_c^0$ as a function of the invasion degree $k_0$ for systems of size $N=1000$ and $\varepsilon=0.05$ on different networks. There has been invasion for the points above the tilted line.}
\end{figure}

The final value of the cooperator density $n_c^{\infty}$ is displayed as a function of $k_0$ for the invasion of a SF network with $N=1000$ agents and $\beta=1.6$ (that is, a  heterogeneous network), and for increasing values of the punishment $\varepsilon$ is shown in Figure~\ref{coopSF}a. All the nodes with degree $k_0$ are set as cooperators at $t = 0$. Then the system is evolved until no more changes are observed in the density of cooperators. Interestingly, in the range of low values of $k_0$,  $n_c^{\infty}$ decreases until it reaches a minimum $n_c^{\min}$, after which it increases and tends to a maximum value $n_c^{\max}$ for a very high $k_0$. This means that the chances of cooperators to invade the network strictly depend on $k_0$. Given the shape of the degree distribution $P(k) \sim k^{-\beta}$, the number of nodes with low degree is higher and a competition effect appears between adding more initial cooperators when $k_0$ is small and the efficiency of the nodes to propagate the cooperative behavior, which seems to be stronger at higher $k_0$ values. This explains the initial decay of $n_c^{\infty}$ and its ulterior strong increase. Different parameters $\varepsilon$ produce some changes in the level of final fraction of cooperators $n_c^{\infty}$. However, the curves of $n_c^{\infty} (k_0) $ follow the same qualitative behavior.

\begin{figure}
\centering
\includegraphics[width=8.6cm]{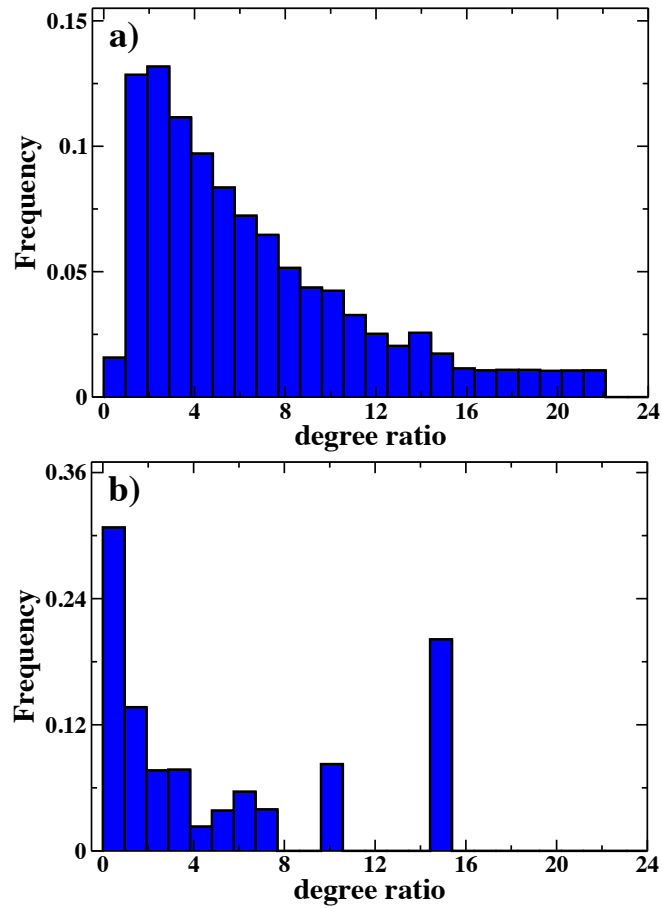}  
\caption{\label{dir} Histograms of the frequency of transitions from defection to cooperation as a function of the ratio $k_i/k_f$, being $k_f$ the degree of the agent which flipped from defection to cooperation by imitating the agent with degree $k_i$, for a system on a scale free network (exponent $\beta=1.6$ and size $N=2000$, $\varepsilon=0.05$ and $k_0=30$), in case of {\bf a)} UI evolution rule, and {\bf b)} REP updating. The cumulative frequency of the transitions with degree ratio larger than one ({\it i.e.} the top-down invasion acts) is $\simeq 98\%$ for UI and $\simeq 69\%$ for REP.}
\end{figure}

When the size of the system is varied for a fixed $\varepsilon$, the picture emerging is similar. In Figure~\ref{coopSF}b, $n_c^{\infty}$ is depicted as a function of $k_0$ for $\varepsilon = 0.05$. The invasion from nodes of degree $k _0 = 2$ have a low ratio of success, which increases to values of the order of one if the initial cooperative nodes are the hubs. The intermediate $k_0$ values lead to a very small $n_c^{\infty}$, which becomes smaller and smaller as the system size increases. 

Similar results are obtained  with a less heterogeneous network ($\beta=2.7$), as shown in Figure~\ref{coopSF}c: the only difference is that after reaching a maximum, $n_c^{max}$ tends to vanish for $k_0\rightarrow\infty$ due to the scarcity of hubs. In the extreme situation, the scenario is modified if we consider almost homogeneous networks as can be seen in Figure~\ref{coopER}a. We show again the behavior of $n_c^{\infty}$ as a
function of $k_0$ for different values of $\varepsilon$, but on an ER network with average
degree $\langle k\rangle=3.5$. In this case, we see a much simpler behavior: the final
cooperation level is always very low, and rapidly decreases with increasing $k_0$.
The same behavior can be observed on a SF network but with $\beta=5.6$,
that is, a network much closer to a homogeneous one than a SF with $\beta < 3$.

\begin{figure}
\centering
\includegraphics[width=8.6cm]{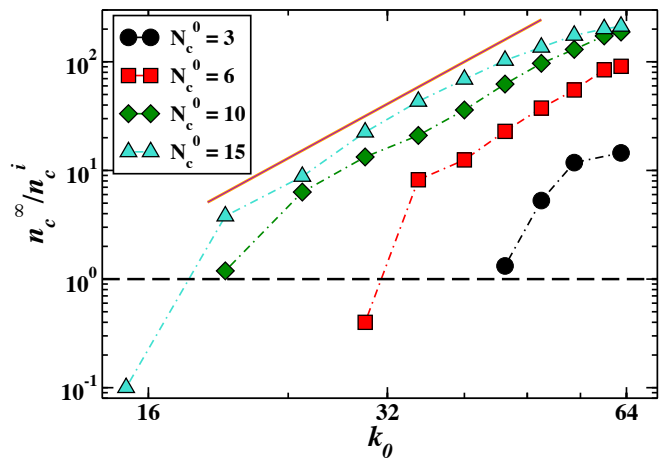}
\caption{\label{fix1} Values of ratio between the final cooperator density $n_c^{\infty}$ over the initial one $n_c^i$ as a function of the invasion degree $k_0$ for systems of size $N=4000$ on an SF network (exponent $\beta=1.6$), $\varepsilon=0.05$ and different number of initial invaders; the lacking points for small invasion degrees are values of $k_0$ for which cooperators end up totally wiped out; the violet straight line represents a power law with exponent $4$. There has been invasion for the points above the tilted line.}
\end{figure}

In all the configurations that we have investigated the final cooperation level never reaches the unit density, so that cooperative behaviors are not able to completely invade the network. Furthermore, the resulting  $n_c^{\infty}$ is averaged over thousands of realizations. In some of them the population of cooperators may have extinguished, providing $n_c^{\infty}$ a clue of the probability of persistence of cooperators at $t \to \infty$. When the invasion properly occurs, the final density of cooperators must grow respect to the initial one. This is why we will talk about a proper invasion when the ratio between the final cooperator density and the initial one is larger than $1$.  Figure~\ref{glob_c} shows this ratio $n_c^\infty/n_c^0$ as a function of the degree $k_0$ for different types of networks. According to these results, a high degree heterogeneity in the network is necessary for cooperative behaviors to invade (we only see it if $\beta <3$). Not only that, it is also required that $k_0$ is over a certain value for cooperation to spread. 

\subsection{Direction of the invasion process }

An important question is through which modality the invasion process takes place (when it does). In particular, when a defector imitates a cooperator, {\it i.e.} a site is invaded by cooperation, it is rather relevant to know if the invaded node has a higher degree than the invader or not.  Figure~\ref{dir}a shows that a single invasion act is more likely to happen top-down than bottom-up, that is, there is a statistical bias that favors configurations with the invaded node having lower degree than the invading one. The distribution of the ratios between the degree of the initial cooperative node $k_i$ and that of the node adopting cooperation $k_f$ is shown in Figure~\ref{dir}a. The first bin, between zero and one, encloses all the instances when $k_f \ge k_i$. As can be seen, this is a much smaller fraction of all the invasion processes registered, and this happens also with a different update rule, as shown in Figure~\ref{dir}b. In principle, this result could be a consequence of the friendship paradox~\cite{fel91}. Anyway, if we compute the average degree ratio $r_d=\langle k_n/k\rangle$ (being $k_n$ the general neighbor's degree) for the network utilized in Figure~\ref{dir} (Molloy-Reed scale free network with $N=2000$ nodes and exponent $\beta=1.6$), it results $r_d\simeq1.59$, whilst the average ratio between the invader degree and the invaded one is around $7$ for the UI dynamics and $5.8$ for the REP rule.

\begin{figure}
\centering
\includegraphics[width=8.6cm]{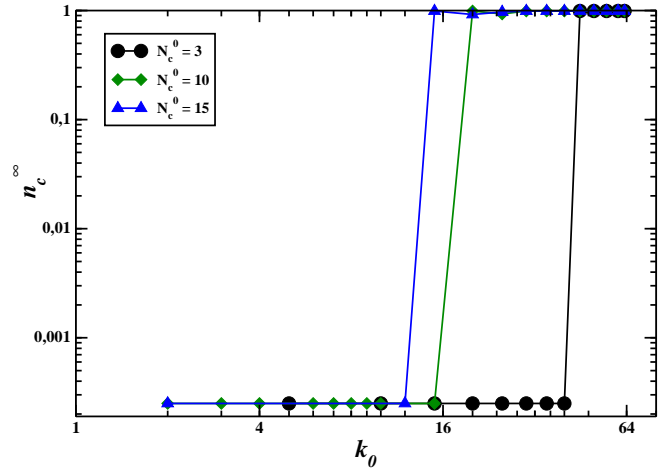}% This is an * file
\caption{\label{fix2} Cooperator density $n_c^{\infty}$ in the last mixed configuration over $n_c^i$ as a function of the invasion degree $k_0$ for systems of size $N=4000$ on an SF network (exponent $\beta=1.6$), $\varepsilon=0.05$ and different number of initial invaders. This stage is always with just one contrarian which will be reabsorbed at the next step of the dynamics.}
\end{figure}

\subsection{ Fixed number of initial invaders }
\label{fixed_n}

In order to better understand the mechanisms of the invasion, we also performed some simulations setting at the initial stage of the dynamics only defectors, apart $N_c^0$ cooperators on nodes of degree $k_0$. Naturally, for small values of $k_0$ the network has always at least $N_c^0$ nodes of such degree, whilst for large $k_0$ we have kept only realizations of the network having the number of nodes needed. This choice allows us to better understand the mechanisms underlying the observed dynamics. In Figure~\ref{fix1}, we see that if the invasion does not take place (that is, if the final cooperator density is not larger than the initial one), then the cooperator density is generally vanishing. This is even more understandable watching Figure~\ref{fix2}, where we show the average cooperator density in the last stage where both strategies are still present in the system. As it can be easily noticed, with fixed number of invaders the final configuration is always all-cooperators or all-defectors, while setting initially all the nodes of grade $k_0$ with invaders entails the possibility of a final mixed state.
As the reader can notice, for these simulations we utilized larger networks, so that the systems can have more nodes with larger degrees than in the previous case, magnifying strength and weakness of high degree invaders.

Such results reinforce the previous considerations: the outcome of the dynamics depends on the combined effect of the degree of the initial invaders and their quantity: without fixed number of initial cooperators (baseline model configuration), increasing $k_0$ is helpful for the invaders because it increases their influence towards the rest of the system, but, at the same time, reduces their number weakening the enhancing effect of a higher degree. Therefore, if we set $N_c^0$, for small $k_0$, we have generally less invaders with respect to the baseline (all the nodes of degree $k_0$ cooperators at $t=0$), whilst for high values of $k_0$ we have even more invaders than in the baseline: in practice, if in the previous case we have always two competing effects (a lot of invaders with little connectivity in one case, few invaders with big connectivity in the second one), by setting $N_c^0$ we have in both cases two adding up effects (few, weak invaders for small $k_0$, and many, powerful invaders for large $k_0$).
To be even clearer, let us suppose that for a given high value of $k_0$ the average number of nodes with such degree be two: in this case, only two initial invaders have a very low probability to survive, but if we require to have 5 or 6 of them, selecting only the few iterations where the network has generated them, we will have a bunch of invaders with high degree and, consequently, very high probability to form a cluster able to completely invade the system, as we will show also in Sec.~\ref{TC}.
As a result, there is always a complete invasion of cooperators (for large $k_0$), or a total extinction (for small $k_0$). In short, we can conclude that in order to allow the cooperation to spread throughout the system, it is important that the initial cooperators are located on nodes of degree large enough to permit the initial cooperative cluster have many links, but not too large to reduce excessively the number of nodes in such cluster and the connections among themselves.

\subsection{Average Payoff}
\label{av_p}

\begin{figure}
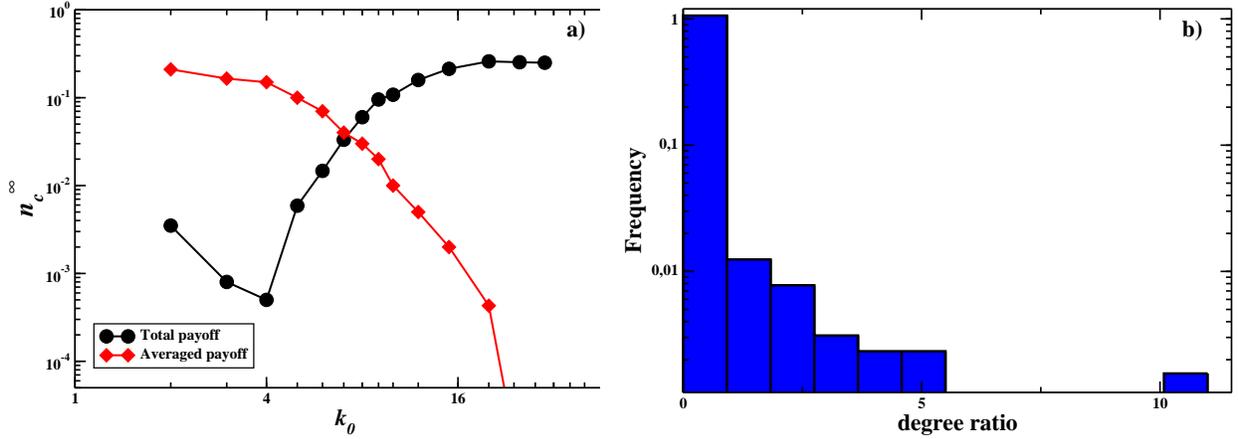
%[h]
\begin{centering}
\includegraphics[width=8.1cm]{Figure7A.eps}
\includegraphics[width=8.1cm]{Figure7B.eps}
\end{centering}
\caption{
{\bf a)} Final cooperator density as a function of the invasion degree $k_0$ for a system of $N=1000$ agents on a scale free network with $\beta=1.6$, $\varepsilon=0.05$; fitness given by the average payoff per neighbor (black), and by the total payoff collected (red, same curve shown in Fig.~\ref{coopSF}a).
{\bf b} Histogram of the frequency of transitions from defection to cooperation as a function of the ratio $k_i/k_f$, for the same system of left panel (average payoff) and $k_0=4$; notice the logarithmic $y$-scale.}
\label{new}
\end{figure}

In this section we aim to evaluate how changing the way agents collect their payoff affects the behavior of the model. In particular, as we anticipated in Sec.~\ref{themodel}, we consider the case in which the payoff collected at each round by an agent is divided by her degree, that is, the fitness is given by the average payoff per neighbor instead that by the total payoff. As illustrated in Fig.~\ref{new}, here the behavior of the model is deeply different from the one previously observed. In particular, we see that the cooperation survives only for very low invasion degrees, then vanishes exponentially fast as $k_0$ increases. Moreover, the few transitions from defection to cooperation prefer clearly the bottom-up direction, differently from the baseline case. In practice, averaging the payoff over the neighbors makes higher degree nodes lose their importance as will be discussed in the final section.

\subsection{Moody conditional cooperation updates}

Up to now, we have shown that the spreading of cooperation is more likely to take place from higher-degree towards lower-degree nodes, setting a preferential direction for the invasion process. We verified this outcome with UI and REP evolution algorithms, but we may also wonder if this effect is more general, and can be detected even when the elementary dynamics is deeply different. In practice, we want to establish if the mechanisms at work in a considerably different model drive the system to the same result. Let us consider, for example, the Moody Conditional Cooperation (MCC) dynamics~\cite{grujic2010}. The moody conditional cooperation was proposed as a probabilistic update rule to explain the decision patterns observed among individuals playing a Prisoner's Dilemma in a large-scale experiment. It is, therefore, closer to real human decision making. The main feature of moody cooperation is not the evolution algorithm, but the definition itself of the strategies for each player. In the model as defined in Sec.~\ref{themodel}, the strategy of an individual is in every moment univocally determined, so that when involved in a game round, her action is already determined. On the contrary, with MCC the action of the players depends on the number of cooperating neighbors
they had in the previous round: the more neighbors cooperated,
the more likely it is that the player cooperates. However, the
choice depends also on player's own previous action: thus,
it has been shown that cooperation following cooperation
is much more likely than following defection~\cite{gracia2012b,gru14}.

\begin{figure}
\centering
\includegraphics[width=8.6cm]{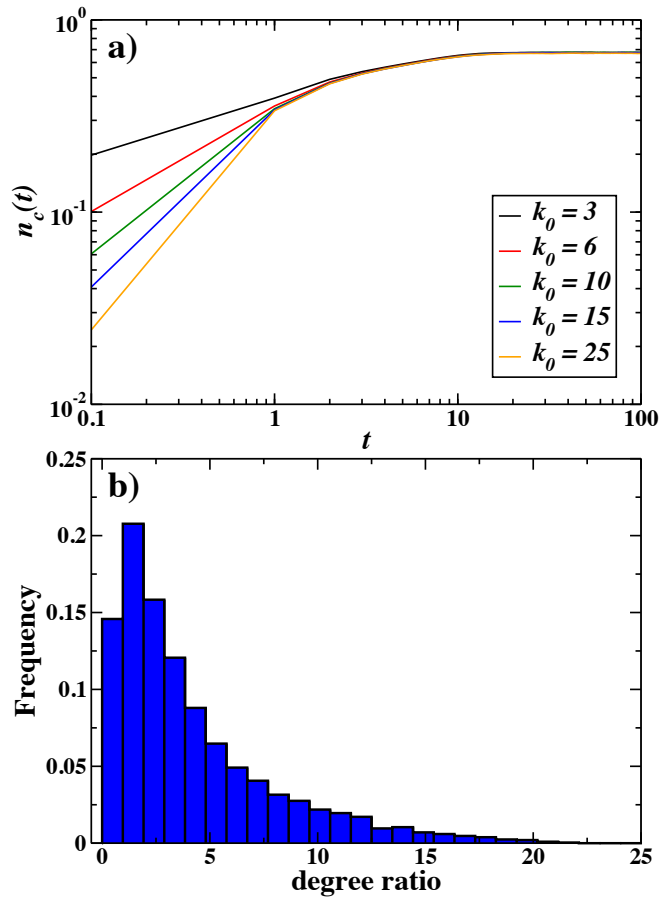}
\caption{\label{mcc} System on a SF network $N=1000,\ \beta=1.6$ with Moody Conditional Cooperation dynamics. In {\bf a)}, time behavior of the cooperator density for different values of the initial invasion degree. In {\bf b)}, histogram of the frequency of transitions from defection to cooperation as a function of the ratio $k_i/k_f$. The cumulative frequency of the transitions with degree ratio larger than one is $\simeq 85 \%$. Considering an ER network leads to very similar results.}
\end{figure}

We implemented the Moody Conditional Cooperation dynamics as follows. The probability $P_c$ to cooperate
of an individual is
\be 
\label{mcc_alg}
P_C=
\left\{
\begin{array}{ll}
pX+r & \ \ \ \ \ \mbox{if cooperated in the previous round}  \\
 & \\
q & \ \ \ \ \ \mbox{if defected in the previous round} ,
\end{array}
\right.
\ee
where $X$ is the fraction of cooperating neighbors in the previous interaction, and $p, q, r \in[0,1]$
are the quantities defining the individual's behavior: in practice the set $\{p^i,q^i,r^i\}$ defines the (complex) strategy of the player $i$ (of course, if it is $pX+r>1$, the probability $P_C$ is set equal to 1). After the interaction, that is, after $i$ and her neighbors have played a game round, each one with her own neighbors, the strategy evolves according the UI rule: if at least one neighbor earned more than herself, $i$ will imitate the best performing one, that is, she will adopt the set $\{p,q,r\}$ of the fittest neighbor.
The strategy parameters are initially distributed at random, and at the first interaction every 
player is considered as a previous defector, apart those occupying nodes of degree $k_0$, which are defined as cooperators.

The Moody Conditional Cooperation dynamics is well known to be little influenced by the topology of the network~\cite{gru14} (we found this same behavior in this case), and proved to be more realistic~\cite{grujic2010}, as it was observed in experiments with human subjects. Besides, since players' strategy is not a defined action (cooperation or defection), but a mixed complex one, it is quite hard to define completely an invasion process. However, it is always possible to study the time evolution of the cooperator density ({\it i.e.}, the fraction of cooperating actions per unit time). Moreover, if we consider an agent which cooperates after having defected at the previous game round, and consider the ratio between the degree of the neighbor she imitated (when the change of action is the actual consequence of a change of strategy) and her own degree, we can draw an histogram as those in Figure~\ref{dir}.

Here we resume the main results for the Moody Conditional Cooperation dynamics in Figure~\ref{mcc}. First of all, the final outcome of the evolution is independent from $k_0$, since in this case also the defectors with no cooperating neighbors have a finite probability to flip action to cooperation. Secondly, considering the transitions from defection to cooperation, when induced by a strategy imitation, we see that also in this case the vast majority of such transitions take place from higher to lower degrees. Finally, we stress the fact that changing the type of network in this case does not change the results obtained with the SF network utilized in Figure~\ref{mcc}. These results allow us to conclude that, also with a totally different dynamics, cooperative behaviors, in a population of individuals interacting as in Prisoner's Dilemma, spread essentially from higher to lower degree nodes (when they do). This means that this kind of process is very general and does not depend strictly on the details of the model, but is quite universal. In fact, since in the Prisoner's Dilemma cooperation is at individual level a disadvantageous behavior, cooperators with higher degree, which in complex networks are also likely to be directly connected, can sustain more easily their pro-social strategy, and therefore contribute efficiently to the invasion of less connected agents.

\ 

\section{Analytical discussion}
\label{TC}

In order to shed light on the fundamental mechanisms which give origin to the phenomenology presented in the previous sections, we have to analyze the actual effect of the topology on the dynamics. In order to do that, in this section we  proceed in two ways. First, we will try to apply to our model a peculiar mean-field approach for networks already utilized in literature for the study of reaction-diffusion processes. Afterwards, to overcome the limits of such treatment, we will consider more qualitatively the effect of the spatial fluctuations through the network on the model dynamics. 

\subsection{Heterogeneous mean-field}

A possible way ahead could be to study the time evolution of the partial cooperator densities $n_c^k(t)$ (that is, the fraction of cooperators occupying nodes of degree $k$), following an already developed approach utilized for reaction-diffusion processes on heterogeneous networks~\cite{cat05}.
We start by defining the single node occupation number $n_i^t$ in this way:
$\nu_i^t=1$ if a cooperator occupies the site $i$ at time $t$, $\nu_i^t=0$ if
instead the node $i$ is occupied by a defector. Its evolution rule is
\be 
\label{app_rule1}
\nu_i^{t+1}=\nu_i^t \, \eta_i+(1-\nu_i^t) \, \xi_i ,
\ee
where $\eta_i$ and $\xi_i$ are quantities given by
\be
\label{app_rates1}
\eta_i=
\left\{
\begin{array}{ll}
0 & \ \ \ \ \ \mbox{with probability \ \ } \lambda_i \\
 & \\
1 & \ \ \ \ \ \mbox{with probability \ \ } 1-\lambda_i
\end{array}
\right.
\ee
and
\be
\label{app_rates2}
\xi_i=
\left\{
\begin{array}{ll}
0 & \ \ \ \ \ \mbox{with probability \ \ } 1-\mu_i \\
 & \\
1 & \ \ \ \ \ \mbox{with probability \ \ } \mu_i \ , 
\end{array}
\right.
\ee
being $\lambda_i$ ($\mu_i$) the probability that a cooperator (defector) in node $i$ becomes a defector (cooperator) after time $t$ (for simplicity, we keep implicit their time dependence).
It is easy at this point to compute the average over the ensemble:
\be 
\label{app_rule2}
\langle \nu_i^{t+1}\rangle=\langle \nu_i^t\rangle+\mu_i-(\lambda_i+\mu_i)\langle \nu_i^t\rangle \ , 
\ee
which, passing to continuous time and defining $n_c^i(t)\equiv\langle\nu_i^t\rangle$, becomes
\be 
\label{app_rule3}
\dot{n}_c^i=\mu_i-(\lambda_i+\mu_i)n_c^i(t) \ . 
\ee
Assuming that the nodes of the same degree are statistically equivalent (uncorrelated network), Equation~(\ref{app_rule3}) can be rewritten as
\be 
\label{app_rulek}
\dot{n}_c^k=\mu_k-(\lambda_k+\mu_k)\, n_c^k(t) \ , 
\ee
where, naturally, the index $k$ refers to all the agents occupying vertices of degree $k$. Now, evaluating the factors $\lambda_k$ and $\mu_k$ is very hard: as pointed out at the beginning of this section, the probability for an agent to switch strategy depends on the distribution of cooperators among the nearest neighbors and next nearest neighbors, with the spatial fluctuation playing a fundamental role. Nevertheless, we can deduce that initially the partial cooperator densities have to increase for $k\neq k_0$, whilst $n_c^{k=k_0}$ decreases. Indeed, initially we have by construction $n_c^k(t)\simeq\delta_{k,k_0}$, so that at the early stages of the dynamics, up to some time $t^*$, it must be
\be
\label{app_res}
\dot{n}_c^k(t\lesssim t^*)\simeq
\left\{
\begin{array}{ll}
\mu_k>0 & \ \ \ \ \ \ \ k\neq k_0 \\
 & \\
-\lambda_{k_0}<0 & \ \ \ \ \ \ \ k=k_0 \ , 
\end{array}
\right.
\ee

\noi giving back that, during the initial phase of the evolution, the cooperator density limited to the nodes of degree $k_0$ has necessarily to decrease, whilst it increases in the remaining nodes.

\begin{figure}
\centering
\includegraphics[width=8.6cm]{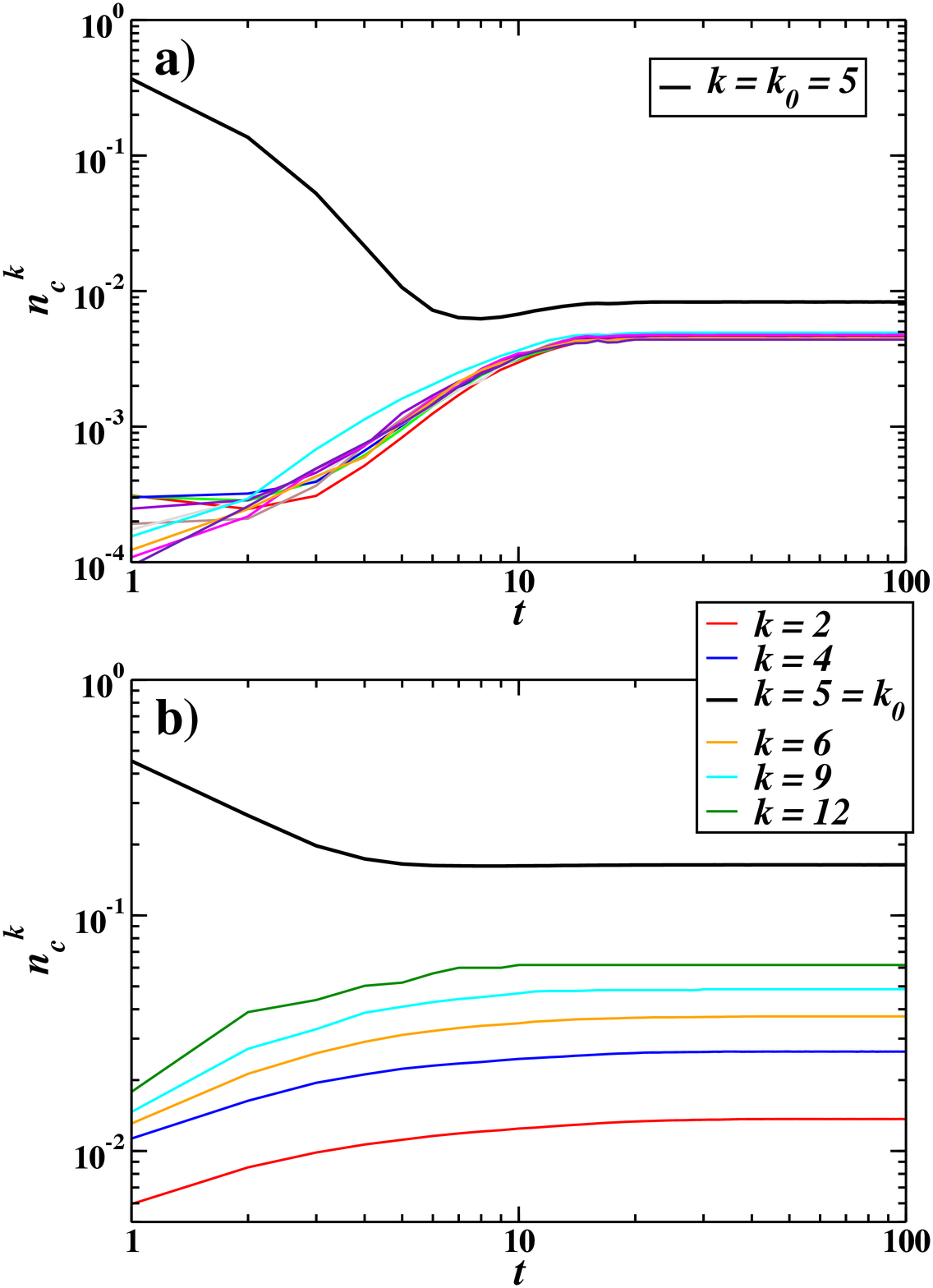}
\caption{\label{kc3} Time behavior of the partial cooperator densities $n_c^k$ for two different topologies. In {\bf a)}, scale-free network (exponent $\beta=1.6$ and size $N=1000$, $\varepsilon=0.05$ and $k_0=5$); the curves shown are for $k$ from 2 to 12 (the partial density for $k=k_0$ is explicitly indicated). In {\bf b)}, ER network (average degree $\langle k\rangle=3.5$ and size $N=1000$), $\varepsilon=0.05$, $k_0=5$, and some values of $k$.}
\end{figure}

The behavior determined by Eq.~(\ref{app_res}) is actually confirmed by simulations (see Figure~\ref{kc3}), but it is not very informative (especially for what concerns the $k_0$-nodes, which initially are all cooperators). Therefore, in order to get more effective information from Equation (\ref{app_rulek}), we should determine the explicit shape of the factors $\lambda_k$ and $\mu_k$, which unfortunately is very hard. More precisely, applying this heterogeneous mean-field approach to our model shows two critical points. First of all, mean-field is rigorously valid in the thermodynamic limit (infinite system size) which has no meaning with heterogeneous networks where the average degree diverges with $N$ increasing: this means that Equation (\ref{app_res}) represents correctly the model behavior only at the early stages of the dynamics, after which the finite size effects are no longer negligible. Secondly, the spatial fluctuations are crucial for the outcome of the dynamics. It is well known indeed that in the Prisoner's Dilemma game cooperators survive in highly connected clusters where they can take advantage of mutual cooperation~\cite{now92}: that is, cooperation will spread starting from a bunch of original cooperators of degree $k_0$ accidentally connected among themselves. Therefore, in order to describe completely the entire dynamics, we have to look more in depth at the topological properties of the network.

\subsection{Initial clusterization effects }

For what stated above, we should expect that the invasion of the cooperation is more effective when starting from nodes
of higher degree, but also numerous enough to create an initial cluster of connected cooperators from which the spreading can originate. In order to find out the strength of the invasion as a function of $k_0$, we have to evaluate the probability to have in a given network a cluster of nodes of the same degree $k'$ connected among themselves. To this aim, let us start from the conditional probability $P(k'|k)$ that a node of degree $k'$
is connected with  a node of degree $k$. In the case of uncorrelated networks, we can assume that~\cite{bog03}
\be 
\label{uncN}
P(k'|k)=\frac{k'P(k')}{\langle k\rangle} \ . 
\ee
From the previous equation it is possible to assess the average number $\mathcal{N}_{k_0}$ of neighbors of a node of degree $k_0$ with the same degree:
\be
\label{nk0}
\mathcal{N}_{k_0}=k_0P(k_0|k_0)=\frac{k_0^2P(k_0)}{\langle k\rangle} ,
\ee
which for a SF network is
\be
\label{nk0B}
\mathcal{N}_{k_0}\propto k_0^{2-\beta} \ . 
\ee

\noi We remind here that the Molloy-Reed algorithm we used to generate our SF networks does not allow nodes of degree larger than $\sqrt{N}$, so that Equation~(\ref{nk0B}) can be considered valid for every node. Interestingly, $\mathcal{N}_{k_0}$ is an increasing function of $k_0$ for $\beta<2$, implying that the higher is the invasion degree, the bigger is the initial cooperator cluster and the final number of cooperators, as confirmed by Figure~\ref{coopSF}. The initial decrease for small values of $k_0$ is a finite size effect, as demonstrated
in Figure~\ref{coopSF}b: in the limit of very high $N$, the final partial densities appear to behave as
\be
\label{lim_pd}
n_c^k(t\rightarrow\infty)=
\left\{
\begin{array}{lr}
0 & \ \ \ \ \ \mbox{if } k \leq k^* \\
 & \\
\mathcal{F}(k;k_0,N) & \ \ \ \ \ \mbox{if } k >  k^* ,
\end{array}
\right.
\ee
where $\mathcal{F}(k;k_0,N)$ is an increasing function of $k$ (dependent also on the parameters $k_0$ and $N$). 

On the other hand, for $\beta>2$ the size of the initial cooperator cluster decreases with $k_0$, so that the invasion probability must vanish for $k_0\rightarrow\infty$, as confirmed in Figure~\ref{coopSF}b, even though for $2<\beta<3$ there is an interval of intermediate values of $k_0$ where the balance between the initial degree of cooperators and the size of the invading cluster is still favorable for the spreading of cooperation (this region disappears for larger values of $\beta$, as shown in Figure~\ref{coopSF}b).

Conversely, for an ER random  network, it is easy to understand that Equation~(\ref{nk0B}) reads
\be
\mathcal{N}_{k_0} \propto  k_0^{2} \, e^{ -\gamma \, k_0} \ , 
\label{nk0C}
\ee
\noi where $\gamma>0$. Therefore, $\mathcal{N}_{k_0}$ decays very rapidly with $k_0$, causing the sharp decay of the invasion probability, in its turn confirmed in Figure~\ref{coopER}a. Such result could have been also predicted simply considering that the average number of sites of higher degree is very small in this kind  of networks, so that there are initially too few cooperators to form a cluster.

\ 

\section{Conclusions and perspectives}

In this paper we have studied the mechanisms through which cooperative strategies, which are disadvantageous for individuals though convenient for the population at a global level, can invade an initially hostile environment. The main result we have obtained is that two factors decide if the initial invaders can succeed: they have to be numerous enough to create a safe cluster from where the invasion can spread, but also have enough connections with the other individuals. In fact, on complex networks, if we put the initial invaders only on nodes of the same grade $k_0$, the first requirement (high connectivity) is satisfied for large $k_0$, whilst the second one (high numerosity) is satisfied at small $k_0$. This means that, in order to have a final configuration favorable to cooperation, the initial condition must be a suitable equilibrium between high degree and high numerosity, which takes place, when possible, at intermediate values of $k_0$. More interestingly, our results demonstrate also that the invasion process, when actually works, is a top-down phenomenon, that is, the cooperators occupy more often nodes of lower degree than the starting ones. This is a rather strong result, since we verified it on different networks and with three different update rules: the deterministic Unconditional Imitation; a rule that allows for the presence of noise, the REP; and the Moody Conditional Cooperation dynamics, that is based on empirical observations and which is deeply different from the Unconditional Imitation and Replicator rules. These interpretations of the results are further confirmed by the simulations accomplished considering the average payoff per neighbor: in that case, indeed, the high degree nodes lose their strength in terms of potential fitness, remaining few in number. As a consequence, any attempt to invade the system starting from nodes of not too low degree fails, and only for very small values of $k_0$ some cooperators can finally survive. Of course, further studies should explore the direction of invasion for other imitation rules (e.g., Moran or Fermi rules, or learning algorithms) and investigate whether there are boundary conditions for the top-down invasion process that we have reported; also adding noise in the initial conditions, for example extracting at random from a given distribution the exact number of invaders, will help to understand the most subtle features of the process.

Such results can have important consequences, both theoretical (for the understanding of the cooperative phenomena in nature) and practical ones (to manage correctly social phenomena).
Whilst the fixation probability of a pro-social mutations is of great relevance~\cite{oht06},  our focus in this work is on how the invasion process takes place. The observed fact that the invasion of cooperative strategies follow a preferred top-down direction rises questions with implications in  social and biological sciences alike: for instance, considering the vaccination campaigns in case of epidemics, the present study suggests that in order to make the pro-social behaviors spread through a skeptic population, it would be better to focus on few, well connected individuals than trying to convince as many individuals as possible, with no regards to their connections, which could turn out to be a waste of time and resources. We stress the fact that, even though it is practically hard for policy makers selecting all the individuals with a given number of links, as we did in our model, the most important message that this work is that pro-social strategies have success when they start from a group of agents which have many connections with the external world (since it is a hierarchical process), but also among themselves, so that they can both spread the cooperative behaviours and resist to outer attacks.

Naturally, future studies are needed, both empirical, to confirm or deny these findings in other contexts, and theoretical: in particular, analyzing models where the initial invaders are selected on different basis ({\it i.e.}, by considering their centrality instead of the degree), or utilizing real networks for the simulations, could shed more light on the details of this class of phenomena.

\ 

\section*{Acknowledgments}

D. V. acknowledges the support from H2020 FETPROACT-GSS CIMPLEX, Grant No. 641191 and from the PROTON project, Grant No. N. 699824. V. C. acknowledges the support from the Dutch Research Organization (NWO) Grant No. 612.001.352. J.J.R. acknowledges funding from the Spanish Ministry of Economy and Competitiveness (MINECO) and FEDER (EU) under the grant ESOTECOS (FIS2015-63628-C2-2-R).

\


\begin{thebibliography}{60}
\bibitem{footn} We recall that when according to the update rule an agent imitates another one with higher fitness, evolutionarily this means that the former has been substituted by the fittest (natural selection). Therefore, no information about the other's payoff has reached the imitating individual.
\end{thebibliography}
\end{document}